\documentclass[twocolumn,prl,showpacs,showkeys,amsmath,amssymb]{revtex4}
\usepackage{graphicx}% Include figure files
\usepackage{dcolumn}% Align table columns on decimal point
\usepackage{bm}% bold math
\usepackage{ulem}% package for underlining
\begin{document}
\title{Glory Oscillations in the Index of Refraction for Matter-Waves}
\author{Tony D.\ Roberts, Alexander D.\ Cronin, David A.\ Kokorowski, David E.\ Pritchard}
\affiliation{Massachusetts Institute of Technology, Cambridge, Massachusetts 02139}
\date{\today}
\begin{abstract}
We have measured the index of refraction for sodium de Broglie waves in gases of Ar, Kr, Xe, and N$_2$ over a wide range of sodium velocities.  We observe glory oscillations---a velocity-dependent oscillation in the forward scattering amplitude.  An atom interferometer was used to observe glory oscillations in the \textit{phase shift} caused by the collision, which are larger than glory oscillations observed in the cross section.  The glory oscillations depend sensitively on the shape of the interatomic potential, allowing us to discriminate among various predictions for these potentials, none of which completely agrees with our measurements.
\end{abstract}
 %other possible pacs 34.90.+q, 39.20.+q
%\keywords{}
\maketitle

A singularity in the classical differential scattering cross section $\frac{d\sigma}{d\Omega} = \frac{b}{\sin\theta} |\frac{d\theta}{db}|^{-1}$ arises when $\theta(b_{\rm glory})=0$ or $\pi$ at finite non-zero impact parameter, $b_{glory}$.  This was named glory scattering by B.\ Cellini who observed the resulting circle of back-scattered light around the shadow of his head on the dewy grass.  An even brighter glory can be observed surrounding the shadow of one's airplane on the clouds of spherical water droplets below \cite{generalglory}.

Glory scattering is not just an optical phenomenon.  Interatomic potentials are attractive at long range but have repulsive cores.  Hence there is an impact parameter just beyond the potential minimum for which $\theta(b_{\rm glory}) = 0$.  Quantum mechanics demands that atom-atom scattering be treated as a wave phenomenon, but when the de Broglie wavelength is much smaller than the potential, the semiclassical approximation can be used \cite{fow59} in which the incident wave is viewed as traveling along classical paths.  There are then two possible paths to scatter into the forward direction---a long range diffractive component from large $b$, and the aforementioned path near the potential minimum.  Waves from these paths contribute to the scattering amplitude in the forward direction, $f(k,\theta = 0)$, for which the partial wave treatment reads:
\begin{equation}
f(k,0) = \sum_{l=0}^{\infty} \frac{2l+1}{2k} \sin 2\delta_l + i \sum_{l=0}^{\infty} \frac{2l+1}{k} \sin^2\delta_l,
\label{eq:aa}
\end{equation}
where the sums are both real.
%\begin{figure}
%\includegraphics{glorypaths8}
%\caption{\label{fig:a} The classical trajectories of atom-atom scattering at various impact parameters, $b$.  The %forward scattered paths (bold) interfere, and their relative phase varies with the incident wavevector, $k$.}
%\end{figure}
% (Fig.~\ref{fig:a})

The behaviors of the real and imaginary parts of $f(k,0)$ are different.  $\textrm{Im}f$ (the second sum in Eq.~\ref{eq:aa}) has a large value owing to the fact that $\sin^2 \delta$ averages to $\frac{1}{2}$ where the scattering is strong ($\delta \gg \pi$).  The contributions from large $l$ and $l_{glory}$ add on top of this large value, making $\textrm{Im}f$ less sensitive to the shape of the potential.  On the other hand, the strong scattering region contributes zero on average to $\textrm{Re}f$, and the long range and glory contributions can either add or subtract, making it easier to discern effects from each region as well as their relative phase.

The ratio,
\begin{equation}
\rho(k) \equiv \frac{\textrm{Re}f(k,0)}{\textrm{Im}f(k,0)},
\label{eq:bb}
\end{equation}
has been shown to give information about the rate of increase of the interatomic potential $V(r)$ for large $r$ independently of the strength of $V(r)$ \cite{ber97}.   Also, oscillations in $\rho$ depend sensitively on the potential near the well bottom.  This sensitivity has motivated several theoretical predictions for $\rho$ based on possible interatomic potentials \cite{adv95,fyk96,cad97,fyk97}.

%Our measurement technique for $\rho(k)$ parallels our previous atom interferometer measurement at fixed $k$ \cite{sce95,ber97}.
We have measured $\rho(k)$ in an atom interferometer using a technique that parallels out previous measurement at fixed $k$ \cite{sce95,ber97}.  $\rho$ is found by measuring $\textrm{Re}f$, which is related to the phase shift caused by the collision, as well as $\textrm{Im}f$, which is related via the optical theorem to the absorption cross section \cite{rothefootnote}.  An atom interferometer determines the phase shift and attenuation of the de Broglie wave in one arm of the interferometer that passes through a ``target gas'' by interfering with the other arm which does not.  Specifically, propagation through a gas of length $L$ modifies the wavefunction by a factor $\exp[i(n-1)k_{lab}L]$ where $n$ is the complex index of refraction, $n = 1 + 2\pi n_d f(k,0)/k_{lab}k$, with $n_d$ the target gas density and $k$ and $k_{lab}$ the Na wavevectors in the center-of-mass and lab frames.

\begin{figure}
\includegraphics{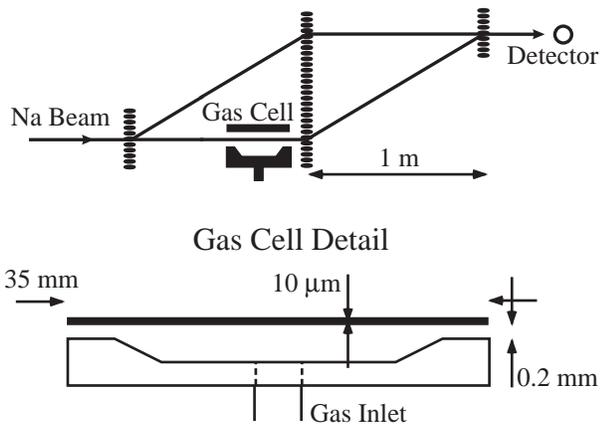}
\caption{\label{fig:b} Schematic of the atom interferometer and the gas cell.}
\end{figure}

The interferometer is composed of three nanofabricated diffraction gratings forming a Mach-Zehnder geometry \cite{ket91}.  A gas cell exposes one path of the interferometer to a gas while leaving the other path undisturbed (Fig.~\ref{fig:b}).  A beam velocity of 0.7--3.0 km/s (with $\sim$5\% rms spread) is chosen by changing the carrier gas mixture in the supersonic oven.  The velocity is measured to $\pm$30 m/s from the atom diffraction pattern of a grating.

Several apparatus improvements were necessary in order to study the velocity-dependent index of refraction.  New 100 nm period gratings \cite{ssc96} double the path separation---and hence the maximum usable velocity in the interferometer---compared to the old 200 nm gratings \cite{rtc95}.  These gratings, developed by Tim Savas and Hank Smith, have already lead to a number of advances in atomic and molecular physics \cite{gratingsfootnote}.
%\cite{anv99,gst99,gst00,hek00}

A thinner gas cell wall \cite{ham97} also allows the atom paths to be closer together and hence allows higher velocities.  The wall of the gas cell is a 10 $\mu$m thick Si wafer which is anodically bonded to a glass substrate.  Channels cut into the glass create the volume of the cell and allow the beam to enter and exit (Fig.~\ref{fig:b}).

The gas cell can be filled and emptied with a time constant of 1--2 sec ($\sim$10 times faster than before) using computer-controlled valves and wider supply lines.  Faster cycle times reduce errors due to phase drift caused by thermal and mechanical motion of the gratings.  This drift has been further reduced by mounting the interferometer on a rigid optical breadboard suspended by a vibration isolation system inside of the vacuum chamber.

$\rho$ is measured by comparing the interference pattern with and without gas in the cell.  In the absence of gas, we observe an interference pattern in the detected signal $I_0$ that depends on the position $x$ of one of the gratings,
\begin{equation}
I_0(x) = N_0 + A_0 \cos (\phi_0 + k_g x),
\label{eq:f}
\end{equation}
where $N_0$ is the average detected flux of atoms (ranging from 5--100 kHz depending on beam velocity), $A_0$ is the amplitude of the interference pattern (typically 5--20\% of $N_0$), and $k_g$ is the grating wavevector.  $\phi_0$, the phase of the interference pattern in the absence of gas, is found by a fit to $I_0(x)$ which also determines $N_0$ and $A_0$.  When the cell is filled with gas, the interference pattern becomes
\begin{equation}
I_{gas}(x) = N_{gas} + A_{gas} \cos (\phi_{gas} + k_g x)
\label{eq:g}
\end{equation}
from which $\rho$ can be determined \cite{sce95}:
\begin{equation}
\rho=\frac{\phi_{gas}-\phi_0}{\ln A_{gas}-\ln A_0}
\label{eq:h}
\end{equation}

A feature of this experimental method is that the measurements of attenuation and phase shift are \textit{relative} measurements, eliminating problems due to intensity fluctuations and phase drift.  Furthermore, measuring the ratio of attenuation vs phase eliminates the dependence on the target gas pressure (which may fluctuate), and obviates the need to know it absolutely.  It also cancels out the $v^{-2/5}$ dependence of cross section and phase shift that can obscure the glory oscillations in $f$.

\begin{figure}
\includegraphics{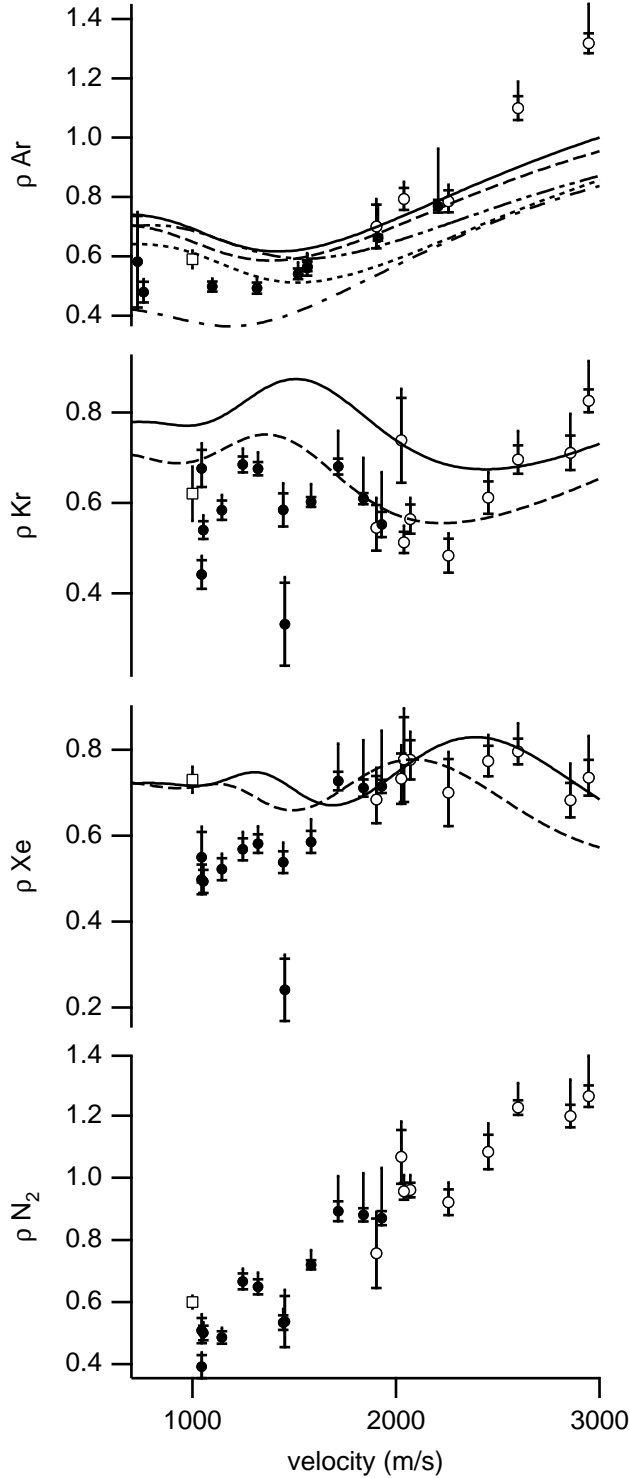}
\caption{ \label{fig:c} $\rho$ as measured for Na waves in Ar, Kr,
Xe, and N$_2$ ($\bullet$ using 200 nm gratings, $\circ$ 100 nm),
showing evidence of glory oscillations in comparison to $\rho$ as
derived from predicted potentials: Na-Ar \cite{zim96}(---),
\cite{dug78}($---$), \cite{fyk97}($\cdots$),
\cite{trk79}($\cdot-\cdot-$),
\cite{tat77}($\cdot\cdot-\cdot\cdot-$); Na-Kr \cite{zim96}(---),
\cite{drs68}($---$); and Na-Xe \cite{bzb92}(---),
\cite{drs68}($---$), \cite{bup68}($\cdots$).  Our 1995 measurement
is also shown \cite{sce95}(boxes).  The horizontal tick marks show
the statistical error.  The vertical error bar line includes
systematic errors.  }
\end{figure}

In Fig.~\ref{fig:c}, $\rho(v)$ is plotted as a function of Na beam velocity $v$ for target gases of Ar, Kr, Xe, and N$_2$.  Each data point represents an average of 1--3 hours of data taken at one to three different pressures at a single beam velocity on a single night.  Our previous index of refraction measurement \cite{sce95} taken at a beam velocity of 1 km/s is shown for comparison.

The improvements to the experimental apparatus mentioned above have allowed us to acquire the large range of data in Fig.~\ref{fig:c} with smaller statistical error than our previous measurements.  We have also been more careful to consider possible systematic errors, correcting for these effects where feasible.

The biggest source of systematic uncertainty is interference from unwanted paths that reach the detector---for instance, the two paths that form the mirror image of the interferometer shown in Fig.~\ref{fig:b}, or from paths of Na$_2$ molecules composing 25\% of the intensity of the beam \cite{ceh95}.  To avoid molecule interference, we deliberately misalign the gas cell wall at low velocities so that it appears wider than the Na$_2$ path separation.  The 50 $\mu$m-wide detector is positioned to avoid the other orders, but because of beam alignment error and thermal drift, there is occasionally some detection of the wrong interference pattern, contributing as much as 5\% to the interfering amplitude, $A_0$, except at higher velocities where it can be 50\%.  The resulting error in $\rho$ is typically $+3$/$-0$\% but as much as $+20$/$-0$\% at the highest velocities of both the 100 and 200 nm data.  This accounts for the asymmetrical systematic error bars in Fig~\ref{fig:c}.

Another systematic effect is the attenuation of the path outside the gas cell due to gas leaking from the ends of the cell.  The integrated density of gas along the path inside the cell relative to the path outside is $120\pm20$ which results in a $+1.7$(0.3)\% correction to the measured $\rho$.

Another source of error arises from measuring the interference pattern while the gas cell pressure hasn't fully equilibrated after filling or evacuating.  We do this to reduce the cycle time to $\sim$20 sec leaving only 3 sec of dead time for filling or emptying.  The resulting error in fitting to the time-averaged interference pattern, up to 0.5\% uncertainty in $\rho$, is included in the systematic error bars.

Contaminants in the target gas must also be avoided.  One such contaminant caused a 3\% uncertainty in $\rho$ for the 200 nm Ar data before it was fixed.

Error bars on the new data points in Fig.~\ref{fig:c} include the sum of statistical and systematic errors.  Because the only non-negligible systematic comes from beam alignment and drift, we have an independent check of this error by examining the repeatability of data taken on different nights.  We expect that the error in alignment or drift may exceed our estimates on some nights, but this can be recognized: $\rho$ will be artificially low compared to other nights, we will see this deviation in all the gases measured, and the measurement will have larger statistical error bars as the ratio changes due to drift.  An example of these features can be seen in the severely outlying data points measured at 1450 m/s for Kr, Xe, and N$_2$.

Another check of systematics can be made by comparing the 100 and 200 nm data where they overlap.  At this velocity, unwanted interference is seven times larger for the 200 nm data than for the 100 nm data, yet the agreement between the two is well within the error estimates.

The data are also consistent with older measurements taken at $v=1$ km/s \cite{sce95}, with the exception of Xe, which disagrees dramatically.  However, no care was taken in that experiment to consider systematic errors, which we estimate were $+30$/$-10$\% in $\rho$ due to the unwanted interference of Na$_2$ molecules and could have been worse for this particular measurement if thermal drift was excessive.

To compare our measurements with a prediction for $V(r)$, the sum in Eq.~(\ref{eq:aa}) must be computed.  The Na wavelength is much smaller than the range of the interatomic potentials we are considering, so hundreds of partial waves contribute to the sum and we are justified in replacing it with an integral:
\begin{subequations}
\label{eq:c}
\begin{equation}
\textrm{Re} f(k) = k \int_0^\infty b \, \sin 2\delta(b) \, db
\label{eq:c1}
\end{equation}
\begin{equation}
\textrm{Im} f(k) = k \int_0^\infty 2b \, \sin^2 \delta(b) \, db
\label{eq:c2}
\end{equation}
\end{subequations}
% (reference as Eq.~(\ref{eq:c1}) and ~(\ref{eq:c2}) or as Eqs.~(\ref{eq:c}))
where $b \equiv (l+\frac{1}{2})/k$ is the classical impact parameter.

Because the kinetic energy is much greater than the well depth of the interatomic potential, we can find $\delta(b)$ using the Eikonal approximation---that $\delta$ is the accumulated phase shift of the atom traveling at constant speed along a straight-line path with impact parameter $b$:
\begin{equation}
\delta(b) = \, -\frac{\mu}{2 k \hbar^2} \int_{-\infty}^\infty dz
\, V(\sqrt{b^2+z^2}),
\label{eq:d}
\end{equation}
where $\mu$ is the reduced mass.  Predictions using this approach are valid to 6\% in $\rho$ for the Na-Ar system and 3\% for the Na-Kr and Na-Xe systems in comparison to an exact quantum treatment \cite{for01}.  Predictions for $\rho$ must also average over the room-temperature distribution of target gas velocities in our experiment, which damps the glory oscillations somewhat at lower beam velocities \cite{cad97,fyk97}.

%Na-Ar \cite{zim96,dug78,fyk97,trk79}; Na-Kr \cite{zim96,drs68}; Na-Xe \cite{bzb92,drs68,bup68}

Fig.~\ref{fig:c} also shows calculations of $\rho(v)$ based on predictions of $V(r)$ for Na-Ar \cite{zim96,dug78,fyk97,trk79,tat77}, Na-Kr \cite{zim96,drs68}, and Na-Xe \cite{bzb92,drs68,bup68} derived from spectroscopic measurements and beam scattering experiments.  The oscillations in these curves are glory oscillations, and are evident in the data.  Even after allowing for statistical and systematic errors, our data disagree with these predictions.  We believe this is because our measurements are very sensitive to the shape of the potential near the minimum (see examples in \cite{rob02}), where the transition from the repulsive core to the Van der Waals potential is poorly understood, even though the well depth and location of the minimum are fairly well known.  The N$_2$ data are remarkable as well for showing a large variation in $\rho$ without any evidence of an oscillation, unfortunately, the potential is not understood in enough detail to explain this behavior \cite{hab80,nsu99}.

Because there is good agreement on the location and depth of the potential minima, we can use this knowledge to scale our measurements of $\rho$ in a way that is sensitive to the \textit{shapes} of all three Na-rare-gas potentials, independent of the well location and depth.  We can write any potential as $V(r)=D_e g(r/r_e)$ where $D_e$ and $r_e$ are the depth and location of the minimum and $g(x)$ is a dimensionless function with a minimum $g(1)\equiv-1$.  In the Eikonal approximation $\rho$ depends only on $g(x)$ and the dimensionless parameter $\alpha \equiv D_e r_e / \hbar v$.  $\alpha$ is approximately the phase shift in radians accumulated along the glory path.  Changing the well depth or location of a potential without changing its shape $g(x)$ is equivalent to changing the velocity scaling of $\rho(v)$.

\begin{figure}
\includegraphics{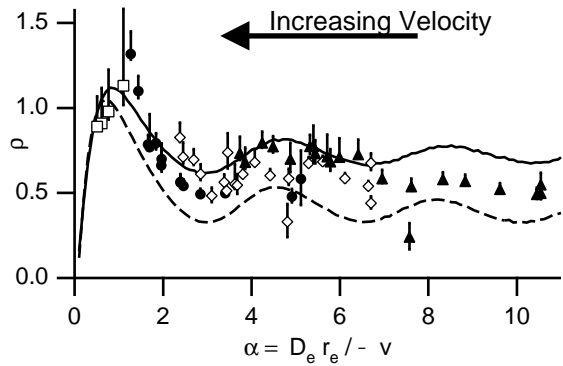}
\caption{\label{fig:d} $\rho$ for the Na-rare-gas systems [Na-Ne
(boxes), Na-Ar (bullets), Na-Kr (diamonds), and Na-Xe (black
triangles)] plotted versus the dimensionless parameter $\alpha
\equiv D_e r_e / \hbar v$.  Also shown is the predicted $\rho$
assuming a Lennard-Jones potential (---), and a Morse potential
($---$).  }
\end{figure}

In Fig.~\ref{fig:d} we have plotted measurements of $\rho$ for Ar, Kr, and Xe verses the dimensionless parameter $\alpha$, using values of $D_e$ and $r_e$ taken from \cite{cad97}.  Because of thermal averaging, the values of $\rho$ we measure also depend weakly on the ratio of $v$ to the mean speed of the scatterer, but differences in $\rho(\alpha)$ due to thermal averaging would be at most of 5\%.  Also included are data for Ne, with new systematic error estimates, which were taken with the older apparatus \cite{sce95,fyk96}.  Plotted in this manner, the combined data clearly show a full glory oscillation, as well as a hint of a second.  For comparison we show $\rho$ as derived from the Lennard-Jones potential, $g(x) = x^{-12} - 2x^{-6}$, and the Morse potential $g(x) = e^{-2\beta(x-1)} - 2e^{-\beta(x-1)}$ using $\beta=4.7$ which is appropriate for the Na-Ar system \cite{fyk97}.  The continuity of the data plotted together in this fashion demonstrates the similarity of Na-rare-gas potentials for these four systems and shows that Lennard-Jones and Morse potentials are too simplistic to accurately model the data here.

In conclusion, we have measured $\rho$ to a typical accuracy of 9\% at various Na velocities in gases of Ar, Kr, Xe, and N$_2$.  The consistency of our scaled data for different gases suggests that the shape $g(x)$ of different Na-rare-gas potentials is similar.  However, discrepancies in the comparison to predictions of $\rho$ suggest that the shape of the potential is not well understood theoretically.  Unfortunately, the difficulty of the inverse scattering problem \cite{new66,chs89} prevents us from deriving the potential based on knowledge of $f(k)$.  However, our explorations of modifications to suggested interatomic potentials do indicate that they would agree better with our measurements if $V(r)$ were modified such that $V(r_0)=0$ occurs at smaller $r_0$ and the well were made less deep in the range $r_e<r<2r_e$ \cite{rob02}.  Our measurements should be useful in evaluating future refinements of the relevant interatomic potentials.

We would like to extend our grateful thanks to Tim Savas and Hank Smith for fabricating gratings for our use.  We thank Edward Smith for the invention of the gas cell, and Peter Finin for his technical assistance.  This work was supported by the Army Research Office Contracts DAAG55-97-1-0236 and DAAG55-98-1-0429, Naval Research Contract N00014-96-1-0432, and National Science Foundation Grant PHY98-77041.  TR acknowledges the support of a National Defense Science and Engineering Graduate Fellowship.

\bibliography{glory}

\end{document}